\documentclass[twocolumn]{openjournal} 

\usepackage{orcidlink} 
\usepackage[T1]{fontenc}
\usepackage{ae,aecompl}
\usepackage[T1]{fontenc}
\usepackage{ae,aecompl}
\usepackage{hyperref}
\usepackage{url}

\usepackage{longtable}
\usepackage{booktabs}
\usepackage{tabu}
\usepackage{graphicx}
\usepackage{multirow}
\usepackage{ulem}
\usepackage{color}
\usepackage{amsmath}

\def \s{~\rm{s}}
\def \km{~\rm{km}}

\def \erg{~\rm{erg}}

\def \yr{~\rm{yr}}

\usepackage{xcolor}
\definecolor{redak}{rgb}{0.9,0.15,0.05}

\shorttitle{Mixing NS material into jets in CEJSNe}
\shortauthors{Soker}

\graphicspath{{./}{figures/}}

\begin{document}

\title{Mass-feeding of jet-launching white dwarfs in grazing and common envelope evolution}

\author{Noam Soker\,\orcidlink{0000-0003-0375-8987}} 
\affiliation{Department of Physics, Technion - Israel Institute of Technology, Haifa, 3200003, Israel;  soker@physics.technion.ac.il}


\begin{abstract}
I propose a scenario that allows white dwarfs (WDs) to launch relatively powerful jets when they enter a common envelope evolution (CEE) or experience a grazing envelope evolution (GEE) with a red giant branch star (RGB) or an asymptotic giant branch (AGB) star. In this, still a speculative scenario, the accretion for a time is mainly onto an accretion disk with a radius of $\simeq 1 R_\odot$ that increases in mass. The accretion disk launches the powerful two opposite jets by releasing gravitational energy, up to several times super-Eddington, as its mass increases. The jets launched by the disk remove high-entropy gas from its outskirts and the envelope that the WD inflates due to nuclear burning on its surface. The motivations to allow WDs to launch powerful jets are recent findings, from the morphologies of post-CEE planetary nebulae, that jets play a major role in the CEE and the accumulating evidence that jets power luminous red novae by jets, as their morphologies indicate. I strengthen my call to include jets in the simulation and modeling of the CEE, consider the GEE as a phase preceding the CEE in many (but not all) cases, and include jets as a major ingredient in modeling and simulating all energetic luminous red novae. 
\end{abstract}

\keywords{stars: jets – stars: AGB and post-AGB – binaries: close – stars: white dwarfs – planetary nebulae: general}

\section{Introduction}
\label{sec:Introduction}

Many planetary nebulae (PNe) morphologies possess a pair or more structural features on opposite sides of the center. The pairs can be of clumps (also called ansae; e.g., \citealt{Balick1987}), lobes, bubbles, and `ears' (for some studies with images of many PNs see, e.g., \citealt{Balick1987, Chuetal1987, Schwarzetal1992, CorradiSchwarz1995, SahaiTrauger1998, Sahaietal2007, Sahaietal2011, Parkeretal2016, Parker2022}). Most studies of the formation of these structures consider shaping by pairs of jets (e.g., \citealt{Morris1987, Soker1990AJ, GarciaSegura1997, SahaiTrauger1998, GarciaSeguraLopez2000, GarciaSeguraetal2005, GarciaSeguraetal2021, GarciaSeguraetal2022, AkashiSoker2018,   EstrellaTrujilloetal2019, Tafoyaetal2019, Balicketal2020, RechyGarciaetal2020, Clairmontetal2022, Danehkar2022, MoragaBaezetal2023, Derlopaetal2024, Mirandaetal2024, Sahaietal2024}; for an alternative scenario see, e.g., \citealt{Baanetal2021}). A large fraction of these studies, as well as other studies of jet-launching by PN progenitors, consider the companion to the asymptotic giant branch (AGB) star to launch the pairs of jets. The companion, a main-sequence star or a white dwarf (WD), accretes material from the AGB envelope, while outside the envelope, or inside it during common envelope evolution (CEE), including grazing envelope evolution (GEE). The same holds for red giant branch (RGB) stellar progenitors of PNe (for PN descendants of RGB stars, see e.g., \citealt{Hillwigetal2017, Sahaietal2017, Jonesetal2020, Jonesetal2022, Jonesetal2023}). 

More than a hundred post-AGB nebulae, pre-PNe, and PNe have a central binary system with a short orbital period, i.e., less than several weeks, indicating that the system underwent a CEE phase (e.g., \citealt{Miszalski2019ic, Oroszetal2019, Jones2020Galax, Jones2025}). Orbital periods of several months might indicate that the system has experienced a GEE. The relation between the post-CEE system and jet-shaped PNe led me to claim \citep{Soker2025Robust} that jets are CEE's most robust observable ingredient. However, the majority of hydrodynamical (and magneto-hydrodynamical) simulations do not include jets that the companion to the AGB or RGB star launches (e.g., a partial list from the last decade \citealt{Staffetal2016MN, Kuruwitaetal2016, Ohlmannetal2016a,  Iaconietal2017b, Chamandyetal2019, LawSmithetal2020, GlanzPerets2021a, GlanzPerets2021b, GonzalezBolivaretal2022, GonzalezBolivaretal2024, Lauetal2022a, Lauetal2022b,  BermudezBustamanteetal2024, Chamandyetal2024, GagnierPejcha2024,  Landrietal2024, RosselliCalderon2024, Vetteretal2024, Vetteretal2025}). Several studies do simulate jets that the more compact companion launches in CEE, including massive stars with a neutron star or a black hole companion; most of these simulations cover a short time, no more than a few orbital periods, or omit other processes (e.g., \citealt{MorenoMendezetal2017, ShiberSoker2018, LopezCamaraetal2019, Shiberetal2019, LopezCamaraetal2020MN, Hilleletal2022, Hilleletal2023, LopezCamaraetal2022, Zouetal2022, Soker2022Rev, Schreieretal2023, Schreieretal2025, Gurjareta2024eas, ShiberIaconi2024}).
These arguments, including the extensive list of CEE simulations, underscore the field's current state: despite considerable progress and efforts over the last decade, we are still far from accurately simulating the CEE, particularly the inclusion of jets that the companion launches. 

In general, the compact companion that enters the envelope of a giant star can be a main-sequence star, a WD, a neutron star, or a black hole. A neutron star or a black hole can accrete at very high rates of  $\dot M_a \gtrsim 10^{-3} M_\odot \yr^{-1}$ and up to $\approx 1 M_\odot \s^{-1}$, because the accreted material can cool by neutrino emission \citep{HouckChevalier1991, chevalier1993, Chevalier2012}; a black hole can as well accrete the gravitational energy. Main-sequence stars can also accrete mass in a CEE. Observationally, some main-sequence stars in post-CEE binary systems of PNe have larger radii than a main-sequence star of the same mass, suggesting they have accreted mass during the CEE (e.g., \citealt{Jonesetal2015}). The large momenta of the nebulae of some PNe (e.g.,  \citealt{Bujarrabaletal2001, Sahaietal2008}) suggest that a main-sequence companion can launch jets in the CEE (e.g., \citealt{BlackmanLucchini2014}). Theoretically, recent studies have demonstrated that if accretion occurs via an accretion disk and the jets launched from the disk remove the outer high-entropy layers of the newly accreted envelope of the mass-accreting star, then the main-sequence star does not expand significantly (\citealt{BearSoker2025acc, Scolnicetal2025}). The main-sequence star accretes at a high rate without much expansion, ensuring that the accreted gas can form an accretion disk around the main-sequence star and that the potential well stays deep. These two effects allow the launching of relatively energetic jets during the CEE, at least the early CEE, and a GEE that might precede the CEE.

This study addresses how a WD might launch energetic jets in a CEE. The WD itself cannot accrete at a high rate because the ignition of nuclear reactions causes the accreted gas to swell. In Section \ref{sec:AccretionRate}, I construct a simple model to estimate the relevant parameters of the accretion process, primarily the accretion rate onto the WD. In Section \ref{sec:Disk}, I describe the accumulation of mass in the unsteady accretion disk and the disk properties. I summarized in Section \ref{sec:Summary} by strengthening the jets as a robust ingredient of CEE and some transient systems, including a fraction of luminous red novae.  

\section{The accretion rate to the WD vicinity}
\label{sec:AccretionRate}

This section does not contain new material but presents the accretion rate with the relevant scaling, considering the negative jet feedback mechanism in the CEE. It is a necessary preparation for Section \ref{sec:Disk}.

For the accuracy of this study, it is adequate to use a simple AGB stellar envelope structure. 
For the density profile of the giant's envelope, I take $\rho_G = \rho_0 (r/100R_\odot)^{-\beta}$. I further scale with an envelope of mass $M_e=1M_\odot$ and an envelope radius of $R_e=200 R_\odot$. Besides the very outer zone just below the photosphere, the profile of extended AGB stars can be approximated with $\beta \simeq 2$. Taking this profile from a very small radio to $r=R_e$, the density profile of the envelope I use here is  
\begin{equation}
\begin{split}
 \rho_G & =  2.35 \times 10^{-7} \left( \frac{r}{100 R_\odot} \right)^{-2}  
 \\ & \times 
 \left( \frac{M_e}{1M_\odot}  \right) 
 \left( \frac{R_e}{200 R_\odot}  \right)^{-1} 
 \quad {\rm for} \quad r<R_e. 
    \label{eq:Density}                            
\end{split}
\end{equation}
The envelope mass inner to radius $r$ is $M_e(r)=(r/R_e)M_e$. For the goal of this study, it is adequate to take the core mass of the AGB star $M_c=0.6 M_\odot$. The relative orbital velocity of the core and the WD during the CEE is 
\begin{equation}
v_o \simeq \sqrt{GM_{\rm in}/a}, 
    \label{eq:Vorb}                            
\end{equation}
 where 
\begin{equation}
M_{\rm in} \equiv M_c+(r/R_e)M_e+M_{\rm WD}, 
    \label{eq:Min}                            
\end{equation}
$a$ is the WD-core orbital separation and $M_{\rm WD}$ is the WD mass. The velocity is smaller because the envelope expands during the CEE, and the mass inside radius $r$ decreases. In scaling equations, I will consider the common WD mass of $M_{\rm WD}({\rm scale})=0.6 M_\odot$ and take $M_{\rm in}({\rm scale}) = 1.5 M_\odot$.   

For the Bondi-Hoyle-Lyttleton (BHL) accretion process in supersonic flows (for the subsonic accretion flow see, e.g., \citealt{Gruzinov2022, Prustetal2024}), the relevant velocity is the relative velocity of the mass-accreting star and the ambient gas $v_r$ (e.g., \citealt{Edgar2004} for a review). 
The BHL accretion radius is defined as 
\begin{equation}
R_{\rm BHL} \equiv \frac{2 G M_{\rm WD} } {v^2_r +C^2_s},  
    \label{eq:RBHL}                            
\end{equation}
where $C_s$ is the sound speed. The relative velocity is $v_r=v_o-v_e$, where $v_e$ is the rotation velocity of the envelope. For the accuracy of the present study, I scale the relations to follow by assuming that the effect of the sound speed in reducing accretion rate and the envelope rotation in increasing it are more or less canceled out by each other. 
I take the accretion rate to be 
\begin{equation}
\dot M_{\rm acc} = \xi \pi R^2_{\rm BHL} \rho v_r.   
    \label{eq:Macc1}                            
\end{equation}
The factor $\xi$ is for an actual smaller accretion rate than the symple formula, and a typical value is $\xi \simeq 0.5$  (e.g., \citealt{Kashietal2022}). 
The density at radius $r$ in the envelope is reduced by the negative feedback of the jets by a factor of $\chi_{\rm j} (NS) \simeq 0.1-0.2$ for a neutron star companion (\citealt{GrichenerCohenSoker2021, Hilleletal2022}), and $\chi_{\rm j} \simeq 0.2-0.6$ for a main sequence companion \citep{WeinerSoker2025}. 

To obtain the approximate accretion rate from equation (\ref{eq:Macc1}), I use the assumptions mentioned above, of taking $v_o$ in the denominator of equation (\ref{eq:RBHL}), $v_r=v_o$, and a density of $\rho=\chi_{\rm j} \rho_G$, yielding  
\begin{equation}
\dot M_{\rm acc} =\pi \xi  \chi_{\rm j} \rho_G   v_o \frac{(2GM_{\rm WD})^2}{v^4_o}. 
    \label{eq:Macc2}                            
\end{equation}
Scaling under these assumptions, I obtain   
\begin{equation}
R_{\rm BHL} \simeq 80 \left( \frac{M_{\rm WD}}{0.4M_{\rm in} } \right)
\left( \frac{a}{100 R_\odot} \right) R_\odot ,
\label{eq:RBHLScale}
\end{equation}
and  
\begin{equation}
\begin{split}
\dot M_{\rm acc} & \simeq 0.19  
\left( \frac{\xi  \chi_{\rm j}}{0.1} \right) 
\left( \frac{M_{\rm WD}}{0.4M_{\rm in} } \right)^2  
\\ & \times
\left( \frac{M_{\rm in}}{1.5M_\odot} \right)^{1/2}                                   
\left( \frac{M_e}{1M_\odot}  \right) 
\\ & \times
\left( \frac{R_e}{200 R_\odot}  \right)^{-1} 
\left( \frac{a}{100 R_\odot} \right)^{-1/2}    M_\odot \yr^{-1} .
\label{eq:MaccScale}
\end{split}
\end{equation}
I note that the accretion radius (equation \ref{eq:RBHLScale}) is of the order of the orbital separation, implying the derived quantities in the present study are accurate up to a factor of a few (half an order of magnitude).  In any case, the accretion rate is $\approx 10 \%$ the BHL rate, and so the actual accretion radius is about a third, i.e., $R_{\rm acc} \approx 25 R_\odot$. 

A WD cannot accrete at such a high rate. A main-sequence star can accrete mass and maintain its identity without expanding much if it ejects the high-entropy gas and loses energy; jets can remove energy and the high-entropy gas and facilitate accretion at high rates (e.g., \citealt{BearSoker2025acc, Scolnicetal2025}). A black hole grows in mass and stays a black hole, and so is a neutron star, as the gas it accretes at high rates cools by neutrino emission (e.g., \citealt{HouckChevalier1991, chevalier1993, Chevalier2012}), until it collapses to a black hole. However, when a WD accretes hydrogen-rich gas at a high rate, it cannot maintain its identity because the rate at which it burns hydrogen to helium is limited (e.g., \citealt{Hachisuetal1999}). The maximum burning rate of a WD of mass $M_{\rm WD} \simeq 0.6 M_\odot$ accreting solar composition gas is $M_{\rm nuc} \simeq {\rm few} \times 10^{-7} M_\odot \yr^{-1}$ (e.g.,  \citealt{Hachisuetal1999}); more massive WDs burn the gas at a higher rate, up to $M_{\rm nuc} \simeq 10^{-6} M_\odot \yr^{-1}$ for a WD close to the Chandrasekhar mass. If the accretion rate is $\dot M_{\rm acc} > \dot M_{\rm nuc}$, the accreting WD turns to a giant as it inflates an envelope (e.g.,  \citealt{Hachisuetal1999}).  
In the next section, I argue that for a limited duration in a CEE, and more so in a GEE, the accretion might be onto an unsteady accretion disk.

\section{Building a non-steady state accretion disk}
\label{sec:Disk}

\begin{figure}
\begin{center}
\includegraphics[trim=3.6cm 14.0cm 0cm 7.0cm, clip, width=0.61\textwidth]
{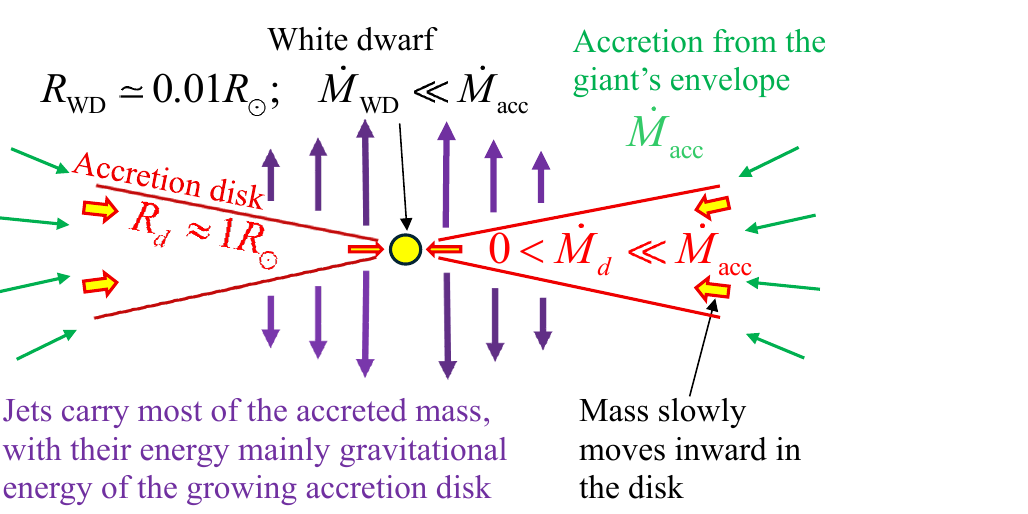}
\caption{A schematic presentation of the proposed scenario (not to scale) in the meridional plane; there is an axial symmetry around an axis through the WD and perpendicular to the accretion disk. The main energy source of the jets is the gravitational energy due to the mass increase of the accretion disk. The WD accretes at a rate below the one that leads to envelope inflation.    }
\label{fig:figure}
\end{center}
\end{figure}

The scenario I propose to allow WDs to launch jets during a GEE and early CEE involves a super-Eddington mass accretion rate onto an unsteady accretion disk at a radius of $R_d \simeq 1 R_\odot$. 
It is this accretion disk that launches the jets. The processes of building an accretion disk at a super-Eddington rate and the launching of a significant fraction of the inflowing gas are similar in some respects to the formation of an accretion disk at a tidal destruction event (e.g., \citealt{Qiaoetal2025}), where the accretion disk is also not in a steady state. 
 Figure \ref{fig:figure} presents a cartoon of the proposed scenario.  

An accretion via a Roche lobe overflow before the onset of the CEE builds the accretion disk around the WD, like in the 3D simulation by \cite{JuarezGarciaetal2025} of pre-CEE mass transfer. \textit{When the WD enters the outskirts of the giant envelope, it already has an accretion disk}. When the WD enters the giant's envelope, the accretion rate increases significantly, and the mass in the accretion disk increases. This occurs in less than an orbital period, as the plunge-in phase is very rapid, occurring over weeks to months (e.g., \citealt{GlanzPerets2021a}). The accretion rate becomes super-Eddington, and the accretion disk launches jets that carry away a large fraction of the inflowing material. 

In \cite{Soker2004}, I discussed a different scenario for the WD to launch jets while in the outer zones of the giant's envelope. In that scenario, the WD inflates an envelope. Due to the angular momentum of the accreted gas, the inflated envelope, of radius $R_s \approx 10 R_\odot$, rapidly rotates, forming a funnel on the two sides along the rotation axis. These funnels blow jets that the nuclear energy of the WD powers. The present flow structure differed in 3 key ingredients.
(1) The accretion is mainly onto a disk, not the WD. (2) The accretion disk, which the rotation supports against the WD gravity, launches the jets rather than an envelope. (3) Gravitational energy powers the jets, not the nuclear burning. The result is that the jets can be more powerful, but operate for a shorter timescale. 

I take the specific angular momentum of the accreted gas as in \cite{Dorietal2023}, which, with the present variables and envelope density profile, reads 
\begin{equation}
j_{\rm acc} \simeq 2 \eta \left( \frac{M_{\rm WD}}{M_{\rm in} } \right)^{3/2} \sqrt{G M_{\rm WD} a},
    \label{eq:Jacc}                            
\end{equation}
where $\eta$ is the ratio of the accreted angular momentum to that entering the BHL accretion cylinder, and it depends on the Mach number and the equation of state; based on \cite{Livioetal1986}, I take  $\eta \simeq 0.25$ for a BHL accretion rate. It is smaller for lower accretion rates, so I scale with $\eta=0.2$.  

The accreted material with this specific angular momentum forms an accretion disk around the WD at a radius of 
\begin{equation}
\begin{split}
R_d \simeq  &  4 \eta^2 \left( \frac{M_{\rm WD}}{M_{\rm in} } \right)^{3} a 
= 1 
\left( \frac{\eta}{0.2}  \right)^2 
\\ & \times
\left( \frac{M_{\rm WD}}{0.4M_{\rm in} } \right)^3  
\left( \frac{a}{100 R_\odot} \right) R_\odot  .
\label{eq:Rdisk}
\end{split}
\end{equation}
Viscosity in the accretion disk spreads the disk inward and outward on a timescale of 10-100 times the Keplerian orbital period of the disk, $\tau_f \simeq 2-20~$day. 

The Eddington limit radiation of a WD of $M_{\rm WD} =0.6 M_\odot$ for electron scattering opacity is $L_{\rm Edd} \simeq 2 \times 10^4 L_\odot$. I scale the accretion onto the disk luminosity (not onto the WD) $L_d=G M_{\rm WD} \dot M_d /2 R_d$, by taking a net accretion rate much lower than the value by equation (\ref{eq:MaccScale}). 
\begin{equation}
\begin{split}
L_d & \simeq 9.4 \times 10^4  
\left( \frac{M_{\rm WD}}{0.6M_\odot} \right)  
\left( \frac{R_d}{1 R_\odot} \right)^{-1}
\\ & \times 
\left( \frac{\dot M_d}{0.01 M_\odot \yr^{-1} } \right) 
 L_\odot  \simeq 5 L_{\rm edd} \approx 5 \dot E_{\rm nuc}.
\label{eq:Ldisk}
\end{split}
\end{equation}
The envelope is optically thick, so the outflow in the jets takes the extra energy. 
The disk's power is larger than the nuclear burning power on the WD, which for a WD of $M_{\rm WD}=0.6 M_\odot$ can reach a value of $L_{\rm nuc} \simeq 1.5 \times 10^4 L_\odot$ \citep{Hachisuetal1999}. Most of the inflowing gas is ejected back. For example, if the inflowing rate is 5 times as large, $\dot M_{\rm inflow} = 5 \dot M_d$, the terminal velocity of the jets, with a mass outflow rate of $\dot M_{r, 2j}=4 \dot M_d$, had they expand freely and not inside the envelope, would be $v_{\rm j} \simeq 0.5 v_{\rm Kep,d} \simeq 170 \km \s^{-1}$, where $v_{\rm Kep,d}$ is the Keplerian velocity in the disk. 
The WD accretes at a much lower rate due to nuclear burning on its surface. For a WD of $M_{\rm WD} =0.6 M_\odot$ the accretion rate is $< 10^{-6} M_\odot \yr^{-1}$ \citep{Hachisuetal1999}. The WD inflates an envelope, but the disk outflow, i.e., the jets, removes the outer zones of the inflated envelope in the scenario I propose. 

At times, the accretion rate onto the disk can be larger than the scaling of equation (\ref{eq:Ldisk}). The proposed scenario suggests that within several months, for large AGB and RGB stars, a WD companion that enters the envelope with an existing accretion disk can accrete at a high rate to increase the mass in the accretion disk around it, even though the WD is unable to accrete at that rate. 
The process itself is limited until a high-pressure state is built from a massive accretion disk, which slows down the accretion rate, or the accretion rate onto the WD itself, $\dot M_{\rm WD}$, surpasses the value above which the WD inflates an envelope. In the latter case, the inflated envelope engulfs the accretion disk and prevents further accretion onto it, shutting down the accretion process via a disk. The disk itself might survive until it is depleted, but no further accretion onto a disk takes place.   

The outflow from the disk removes the high-entropy gas, leaving a low-entropy gas that maintains and builds the accretion disk. The building process of the disk of an average radius of $R_d=1 R_\odot$ releases an energy of 
\begin{equation}
\begin{split}
E_d  \simeq   10^{46} 
\left( \frac{M_{\rm WD}}{0.6M_\odot} \right)  
\left( \frac{R_d}{1 R_\odot} \right)^{-1}
\left( \frac{M_d}{0.01 M_\odot} \right) 
\erg .
\label{eq:Edisk}
\end{split}
\end{equation}
For an escape velocity of $\simeq 50 \km \s^{-1}$ from the surface of the AGB or RGB star, this buildup of the disk can facilitate the removal of $\simeq 0.4 M_\odot$; this represents a significant portion of the outer envelope in the model I consider here.  

Above, I scaled the mass accretion rate onto the disk by $\dot M_d = 0.01 M_\odot \yr^{-1}$, which is $\simeq 5 \%$ of the mass accretion rate in equation (\ref{eq:MaccScale}). The mass accretion might be higher by a factor of a few, and so the accretion power onto the disk might be larger, and within a year, the disk mass can be $\approx 0.03 M_\odot$. If the jets efficiently remove the outer zones of the giant's envelope, the system performs the GEE. The disk expands inward towards the WD, a radius of $\simeq 0.01 R_\odot$, and the accretion releases more energy. The outcome is a transient event with a bright phase that lasts for weeks to a few years, ejecting mass in a bipolar morphology, and the total event energy is $E_{\rm ILOT} \simeq 10^{46} \erg - 10^{47} \erg$. 
\cite{SokerKashi2012} proposed that some progenitors of bipolar PNe experienced an intermediate-luminosity optical transient (ILOT) phase of total (kinetic and radiation) energy of  $E_{\rm ILOT} \simeq 10^{46} \erg - 10^{47} \erg$ and lasted for a time of between one month and three years. The scenario I propose here might lead to similar ILOTs by a binary system of a WD with an AGB or RGB star, which end as bipolar or elliptical PNe.   
  
\section{Summary}
\label{sec:Summary}

This study examined a scenario that would allow a WD companion to an RGB or AGB star to launch energetic jets when it enters the common envelope and spirals inwards in the outer zones of the envelope. The WD itself is limited to accreting mass at a low rate because the nuclear burning of the accreted hydrogen-rich gas inflates an envelope (e.g., \citealt{Hachisuetal1999}). The limit is more than five orders of magnitude below the possible mass accretion rate from the giant envelope as given by equation (\ref{eq:MaccScale}). Instead of accretion onto the WD, I propose that for a time, the accretion occurs onto the accretion disk, which then blows most of the inflowing gas in a bipolar wind, forming the jets. Still, the accretion disk mass grows and releases gravitational energy during that period (Figure \ref{fig:figure}). For typical parameters, the accreted gas from the envelope forms an accretion disk with a typical radius of $R_d \simeq 1 R_\odot$ (equation \ref{eq:Rdisk}). 
Even if accreting at a net rate of $\simeq 10\%$ of the accretion rate according to equation (\ref{eq:MaccScale}), the accretion process onto the accretion disk can liberate $E_d \simeq 10^{46} \erg$ (equation \ref{eq:Edisk}) and few times more within several months to a few years;  this is several times the nuclear burning power of the WD. The accretion process can last longer if the WD performs the GEE, namely, the jets efficiently remove the outer zones of the giant's envelope (the envelope of the RGB or AGB star), and the WD spends some time grazing the giant. In the proposed scenario, the WD already has an accretion disk when it enters the giant's envelope. 

Some earlier studies (e.g., \citealt{Soker2020Galax}) considered the limit of mass accretion onto the WD to be the accretion rate when the nuclear reactions inflate an envelope, $\approx 10^{-6} M_\odot \yr^{-1}$ (e.g., \citealt{Hachisuetal1999}). \cite{BlackmanLucchini2014} find the accretion rate they require for their modeling of momentum deposition by a WD in shaping high-momentum PNe to be super-Eddington. The scenario I proposed here allows super-Eddington accretion into the accretion disk because the jets, rather than radiation, carry away a large fraction of the excess energy that the accretion process liberates. Jets prevent accretion along the polar directions, but have a smaller influence on the accretion from the equatorial plane. 

The accretion scenario I proposed here has speculative parts, like the assumption that the disk can grow in mass (up to a limit) by a super-Eddington accretion (by a factor of several). Numerical simulations of accretion by a WD in a CEE or GEE must confirm this assumed flow structure. As well, the simulations should include the effect of magnetic fields, as the flow develops a dynamo that amplifies magnetic fields (e.g., \citealt{Nordhausetal2011}). My motivations to assume/speculate this flow structure that allows WDs to launch energetic jets at the onset of the CEE, or during a GEE phase that might precede the CEE phase, are arguments that support the major role of jets in the CEE (e.g., \citealt{Soker2025Robust}), and the accumulating evidence for the powering of ILOTs, including luminous red novae, by jets. There are common processes of jets powering transient events (ILOTs, including luminous red novae) and the shaping of PNe (e.g., \citealt{SokerKashi2012, Soker2018Galax}). The morphologies of spatially resolved ILOTs show bipolar morphologies that only jets can explain without difficulties; the list of bipolar luminous red novae includes V4332~Sgr \citep{Kaminskietal2018V4332}, Nova~1670 (\citealt{Sharaetal1985, Kaminskietal2020Nova1670, Kaminskietal2021CKVul, Kaminski2024}), and V838 Mon (\citealt{Chesneauetal2014, Kaminskietal2021V838Mon, Mobeenetal2021,Mobeenetal2024, Mobeenetal2025}). 
Also, jets are much more efficient in converting kinetic energy to radiation in ILOTs, like luminous red novae, than equatorial mass loss or merger without jets (e.g., \citealt{Soker2024Galax}).  
If the accretion process I propose works, a WD can perform the GEE and form jet-powered ILOTs. I argue that most energetic ILOTs, including luminous red novae, are powered by jets, including those ILOTs that studies attribute to CEE or the onset of a CEE (e.g., \citealt{Karambelkaretal2025} for a recent event). Namely, I claim that the launching of energetic jets accompanies the onset of the CEE or the GEE.  

A key process for building a relatively massive accretion disk, $M_d \approx 0.01 M_\odot$, is that the jets that the disk launches remove high-entropy gas from the disk's outskirts and the envelope that the WD  inflates. This is the positive component of the jet feedback mechanism. \cite{Chamandyetal2018} termed this  `pressure release valve', and argued that if it is efficient, super-Eddington accretion can occur. Removing high-entropy gas from an inflated envelope by the jets also allows main sequence stars to accrete at high rates \citep{BearSoker2025acc, Scolnicetal2025}. Overall, the positive component of the jet feedback mechanism, the pressure release valve mechanism, allows main sequence stars and WDs to launch jets in the GEE if it takes place, and the early CEE phase (and possibly during the exit from the CEE phase; e.g., \citealt{Soker2017exit, Soker2019exit}).  

Following this study, I strengthen my call to include jets in the simulation and modeling of the CEE, consider the GEE as a phase preceding the CEE in many (but not all) cases and sometimes replacing the CEE, and include jets as a major ingredient in modeling and simulating all bright luminous red novae, major eruptions of luminous blue variables, and similar gravitationally powered transients, i.e., ILOTs. 

\section*{Acknowledgements}

 I thank an anonymous referee for useful suggestions. 
I thank the Charles Wolfson Academic Chair at the Technion for the support.
A grant from the Pazy Foundation supported this research.




\end{document}